\documentstyle[amstex,amsfonts,12pt]{article}
\catcode`\@=11
\def\numberbysection{\@addtoreset{equation}{section}
        \def\theequation{\thesection.\arabic{equation}}}
\numberbysection
\begin{document}

\newlength{\lno} \lno0.5cm \newlength{\len} \len=\textwidth%
\addtolength{\len}{-\lno}

\setcounter{page}{0}

\baselineskip7mm \newpage \setcounter{page}{0}

\begin{titlepage}     
\vspace{1.5cm}
\begin{center}
{\Large\bf  Reflection matrices for the $U_{q}[sl(m|n)^{(1)}] $ vertex model  }\\
\vspace{1cm}
{\large A. Lima-Santos } \\
\vspace{1cm}
{\large \em Universidade Federal de S\~ao Carlos, Departamento de F\'{\i}sica \\
Caixa Postal 676, CEP 13569-905~~S\~ao Carlos, Brasil}\\
\end{center}
\vspace{2.5cm}

\begin{abstract}
We investigate the possible regular solutions of the  boundary Yang-Baxter
equation for the  vertex models associated with the graded version of the  $A_{n-1}^{(1)}$ affine Lie algebra,
 the $U_{q}[sl(m|n)^{(1)}]$ vertex model, also known as Perk-Schultz model.
 \end{abstract}
\vfill
\begin{center}
\small{\today}
\end{center}
\end{titlepage}

\baselineskip6mm

\newpage

\section{{}Introduction}

The present work is the fourth of \ the four papers devoted to the
classification of the integrable reflection K-matrices for the vertex models
associated with superalgebras. We already have considered the vertex models
associated with the $U_{q}[sl(r|2m)^{(2)}]$ \cite{LIMG}, $%
U_{q}[osp(r|2m)^{(1)}]$ \cite{LIMS1} and $U_{q}[spo(2n|2m)]$\ \cite{LIMS2}
superalgebras. In this paper we have presented the general set of regular
solutions of the graded reflection equation for the $U_{q}[sl(m|n)^{(1)}]$\
vertex model.

Our findings can be summarized into four types of solutions: diagonal
solutions with one free parameter, quasi-diagonal solutions with only two
non-diagonal entries and three free parameters, quasi-diagonal solution with 
$2+2\alpha $ non-diagonal entries in the same secondary diagonal and $%
3+\alpha $ free parameters and, one special type of quasi-diagonal solutions
with $4+2\alpha +2\beta $\ non-diagonal entries in two secondary diagonals
and with $4+\alpha +\beta $ free parameters.

This paper is organized as follows. In the next section we present the $R$%
-matrix of the $U_{q}[sl(m|n)^{(1)}]$ vertex model in terms of standard Weyl
matrices. In the section $3$ we present the solutions of the reflection
equations. In that way we hope that they are the most general set of $K$%
-matrices for the vertex model here considered. Concluding remarks are
discussed in the section $4$. The models with the first values of $m$ and $n$
have its $K$-matrices written explicitly in appendix.

\section{The $U_{q}[sl(m|n)^{(1)}]$ reflection equations}

The $R$-matrix associated with the $U_{q}[sl(m|n)^{(1)}]$ superalgebra \cite%
{Kac, Frappat, Chaichian}, whose matrix elements are the statistical weights
of the Perk-Schultz vertex model \cite{Perk} has the form%
\begin{eqnarray}
R(x) &=&\sum_{i=1}^{N}(-1)^{p_{i}}a_{i}(x)\ {\rm E}_{ii}\otimes {\rm E}%
_{ii}+b(x)\sum_{i,j=1}^{N}\ {\rm E}_{ii}\otimes {\rm E}_{jj}  \nonumber \\
&&+c_{2}(x)\sum_{i<j}^{N}(-1)^{p_{i}p_{j}}{\rm E}_{ji}\otimes {\rm E}%
_{ij}+c_{1}(x)\sum_{i>j}^{N}(-1)^{p_{i}p_{j}}{\rm E}_{ji}\otimes {\rm E}_{ij}
\label{Rsl1}
\end{eqnarray}%
where $N=n+m$ is the dimension of the graded space with $n$ fermionic and $m$
bosonic degree of freedom and ${\rm E}_{ij}$ refers to the $N$ by $N$ Weyl
matrix with only one non-null entry with value $1$ in row $i$ and column $j$.

In what follows we shall adopt the grading structure 
\begin{equation}
p_{i}=\left\{ 
\begin{array}{c}
0,\quad \qquad \quad i=1,2,...,m \\ 
\\ 
1,\quad 
%TCIMACRO{\TeXButton{LaTeX}{\hfill}}%
%BeginExpansion
\hfill%
%EndExpansion
i=m+1,...,N%
\end{array}%
\right.  \label{re.2}
\end{equation}%
and the corresponding Boltzmann weights with functional dependence on the
spectral parameter $u=\ln x$ are given by%
\begin{eqnarray}
a_{i}(x) &=&(x^{(1-p_{i})}-q^{2}x^{p_{i}}),\quad b(x)=q(x-1),  \nonumber \\
\quad c_{1}(x) &=&(1-q^{2}),\quad c_{2}(x)=x(1-q^{2}).  \label{re.3}
\end{eqnarray}%
Here $q$ denotes an arbitrary parameter.

The R-matrix (\ref{Rsl1}) satisfies symmetry relations, besides the standard
properties of regularity and unitarity, namely:

\begin{itemize}
\item {\rm PT} invariance%
\begin{equation}
P_{12}R_{12}(x)P_{12}\equiv R_{21}(x)=R_{12}(x)^{{\rm st}_{1}{\rm st}_{2}}
\label{re.4}
\end{equation}

\item Weaker property \cite{RS, Yue1}:%
\begin{equation}
\left\{ \left\{ \left\{ R_{12}(x)^{{\rm st}_{2}}\right\} ^{-1}\right\} ^{%
{\rm st}_{2}}\right\} ^{-1}=\frac{\zeta (x)}{\zeta (x^{-1}\eta ^{-1})}%
M_{2}R_{12}(x^{-1}\eta ^{-1})M_{2}^{-1},
\end{equation}%
where $\zeta (x)=a_{1}(x)a_{1}(x^{-1})$ and $M$ is a symmetry of the $R$%
-matrix 
\begin{equation}
\left[ R(u),M\otimes M\right] =0,\quad M_{ij}=\delta
_{ij}(-1)^{p_{i}}q^{n+m+1-2i},\quad \eta =q^{n+m}.  \label{re.6}
\end{equation}
\end{itemize}

The matrix $K_{-}(u)$ satisfies the left boundary Yang-Baxter equation \cite%
{Cherednik}, also known as the reflection equation \cite{Sklyanin},%
\begin{equation}
R_{12}(x/y)K_{1}^{-}(x)R_{21}(xy)K_{2}^{-}(y)=K_{2}^{-}(y)R_{12}(xy)K_{1}^{-}(x)R_{21}(x/y),
\label{re.7}
\end{equation}%
which governs the integrability at boundary for a given bulk theory. A
similar equation should also hold for the matrix $K_{+}(u)$ at the opposite
boundary. However, one can see from \cite{Nepo2} that the corresponding
quantity 
\begin{equation}
K^{+}(x)=K^{-}(x^{-1}\eta ^{-1})^{s{\rm t}}M,  \label{re.8}
\end{equation}%
satisfies the right boundary Yang-Baxter equation. Here {\rm st}$={\rm st}%
_{1}{\rm st}_{2}$ and {\rm st}$_{i}$ stands for transposition taken in the $%
i^{th}$ superspace.

Therefore, we can start for searching the matrices $K^{-}(x)$. In this paper
only regular solutions will be considered. Regular solutions mean that the
matrix $K^{-}(x)$ has the form 
\begin{equation}
K^{-}(x)=\sum_{i,j=1}^{N}k_{i,j}(x)\ {\rm E}_{ij}  \label{re.9}
\end{equation}%
and satisfies the condition 
\begin{equation}
k_{i,j}(1)=\delta _{i,j},\quad \qquad i,j=\{1,2,...,N\}.  \label{re.10}
\end{equation}

Substituting (\ref{Rsl1}) and (\ref{re.9}) into (\ref{re.7}), we will get $%
N^{4}$ functional equations for the $k_{ij}$ matrix elements, many of which
are dependent. In order to solve them, we shall proceed in the following
way. First we consider the $(i,j)$ component of the matrix equation (\ref%
{re.7}). By differentiating it with respect to $y$ and taking $y=1$, we get
algebraic equations involving the single variable $x$ and $N^{2}$ parameters 
\begin{equation}
\beta _{i,j}=\frac{dk_{i,j}(y)}{dy}|_{y=1}\qquad i,j=1,2,...,N.
\label{re.11}
\end{equation}%
Second, these algebraic equations are denoted by $E[i,j]=0$ and collected
into blocks $B[i,j]$ , $i=1,...,M-1-i$ and $\ j=i,i+1,...,M-1-i$, defined by 
\begin{equation}
B[i,j]=\left\{ 
\begin{array}{c}
E[i,j]=0,\quad E[j,i]=0,\  \\ 
E[M-i,M-j]=0,\quad E[M-j,M-i]=0.%
\end{array}%
\right. \   \label{re.12}
\end{equation}%
where $M=N^{2}+1.$

For a given block $B[i,j]$, the equation $E[M-i,M-j]=0$ can be obtained from
the equation $E[i,j]=0$ by interchanging 
\begin{equation}
k_{i,j}\longleftrightarrow k_{N+1-i,N+1-j},\quad \beta
_{i,j}\longleftrightarrow \beta _{N+1-i\ ,N+1-j},\quad
c_{1}(x)\leftrightarrow c_{2}(x)\quad \   \label{re.13}
\end{equation}%
and the equation $E[j,i]=0$ is obtained from the equation $E[i,j]=0$ by the
interchanging 
\begin{equation}
k_{i,j}\longleftrightarrow k_{j,i},\quad \beta _{i,j}\longleftrightarrow
\beta _{j,i}\quad  \label{re.14}
\end{equation}%
In this way, we can control all equations and a particular solution is
simultaneously connected with at least four equations.

\section{The $U_{q}[sl(m|n)^{(1)}]$ K-matrix solutions}

Analyzing the $U_{q}[sl(m|n)^{(1)}]\ $reflection equations\ one can see that
they possess a very special structure. The simplest equations are 
\begin{equation}
b(x)\beta _{i,j}k_{i,j}(x)(a_{i}(x)-a_{j}(x))=0\qquad (i\neq j).
\label{bf.1}
\end{equation}%
From (\ref{re.3}), $a_{i}(x)\neq a_{j}(x)$ when the labels $i$ and $j$ are
different types of degree of freedom. It means that all $%
U_{q}[sl(m|n)^{(1)}] $ reflection matrices have the following block diagonal
structure

\begin{equation}
K^{-}(x)=\left( 
\begin{array}{cc}
K^{b} & {\rm 0}_{m\times n} \\ 
{\rm 0}_{n\times m} & K^{f}%
\end{array}%
\right)  \label{bf.2}
\end{equation}%
where $K^{b}$ is a $m$ by $m$ matrix with entries $k_{i,j}$ for $%
i,j=\{1,2,...,m\}$ and $K^{f}$ is a $n$ by $n$ matrix with entries $k_{r,s}$
for $r,s=\{m+1,m+2,...,N\}$.

Now, by direct inspection of the equations (\ref{re.12}), one can see that
the diagonal equations $B[i,i]$ are uniquely solved by the relations%
\begin{equation}
\beta _{i,j}k_{j,i}(x)=\beta _{j,i}k_{i,j}(x),\qquad \forall \ i\neq j.
\label{sol.1}
\end{equation}%
It means that we only need to find the $m(m-1)/2$ and $n(n-1)/2$ elements $%
k_{i,j}$ with $i<j$. \ Now we choose a particular $k_{i,j}$ $(i<j)$ to be
different from zero, with $\beta _{i,j}\neq 0$, and try to express all
remaining non-diagonal matrix elements in terms of this particular element.
We have verified that this is possible provided that 
\begin{equation}
k_{r,s}(x)=\left\{ 
\begin{array}{c}
x\frac{\beta _{r,s}}{\beta _{i,j}}k_{i,j}(x)\quad {\rm if}\quad r>i\quad 
{\rm and}\quad s>j \\ 
\\ 
\frac{\beta r,s}{\beta _{i,j}}k_{i,j}(x)\quad {\rm if}\quad r>i\quad {\rm and%
}\quad s<j%
\end{array}%
\right. ,\qquad (r\neq s)  \label{sol.2}
\end{equation}%
Combining (\ref{sol.1}) with (\ref{sol.2}) we will obtain a very strong
entail for the elements out of the diagonal 
\begin{equation}
k_{i,j}(x)\neq 0\Rightarrow \left\{ 
\begin{array}{c}
k_{p,j}(x)=0\quad {\rm for}\quad p\neq i \\ 
\\ 
k_{i,q}(x)=0\quad {\rm for}\quad q\neq j%
\end{array}%
\right.  \label{sol.3}
\end{equation}%
It means that for a given $k_{i,j}(x)\neq 0$, the only elements different
from zero \ in the $i^{th}$-row and in the $j^{th}$-column are $%
k_{i,i}(x),k_{j,i}(x),k_{j,j}(x)$.

Analyzing more carefully these equations with the conditions (\ref{sol.1})
and (\ref{sol.3}), we have found from the $m(m-1)/2$ elements $k_{i,j}\
(x)(i<j)\in K^{b}$ \ and $n(n-1)/2$ elements $k_{i,j}\ (x)(i<j)\in K^{f}$
that there are three possibilities to choose a particular $k_{i,j}(x)\neq 0$:

\begin{itemize}
\item Only one non-diagonal element and its symmetric are allowed to be
different from zero. Thus, we have $m(m-1)/2$ reflection $K$-matrices with $%
N+2$ non-zero elements and $n(n-1)/2$ reflection $K$-matrices with $N+2$
non-zero elements. These solutions will be denoted by ${\Bbb K}_{[ij]}^{(0)}$
and named {\rm Type-I} solutions.

\item For each $k_{i,j}(x)\neq 0,$ additional non-diagonal elements and its
asymmetric are allowed to be different from zero provided they satisfy the
equations 
\begin{eqnarray}
k_{i,j}(x)k_{j,i}(x) &=&k_{r,s}(x)k_{s,r}(x),  \nonumber \\
i+j &=&r+s\quad {\rm with\quad }\{i,j,r,s\}\in K^{b}\text{ {\rm or} }K^{f} 
\nonumber \\
&&  \label{rule2}
\end{eqnarray}%
It means that we will get a $K$-matrix with entries of the diagonal
principal and the entries of a diagonal secondary with the element $%
k_{i,j}(x)$ on the top. These solutions will be denoted by ${\Bbb K}%
_{[ij]}^{(\alpha )}$ and named {\rm Type-II} solutions.

\item For each $k_{i,j}(x)\neq 0,$ additional non-diagonal elements and its
asymmetric are allowed to be different from zero provided they satisfy the
equations 
\begin{eqnarray}
k_{i,j}(x)k_{j,i}(x) &=&k_{r,s}(x)k_{s,r}(x),  \nonumber \\
i+j &=&r+s\quad\mod{N}\quad {\rm with\quad }\{i,j\}\in K^{b}\text{ {\rm and} 
}\{r,s\}\in K^{f}  \nonumber \\
&&  \label{rule3}
\end{eqnarray}%
It means that we will get a $K$-matrix with the diagonal principal elements
and the elements of two diagonal secondary with the top elements $%
k_{i,j}(x)\in K^{b}$ and $k_{r,s}(x)\in K^{f}$. These solutions will be
denoted by ${\Bbb K}_{[ij][rs]}^{(\alpha )(\beta )}$ and named {\rm Type-III}
solutions.

Here the symbols $\alpha $ $\ $and $\beta $ mean the number of additional
pairs of non-zero entries ($k_{a,b}(x),k_{b,a}(x)$) on the secondary
diagonals.
\end{itemize}

\bigskip For example, the $U_{q}[sl(4|2)^{(1)}]$ model has the following $K$%
-matrix 
\[
K=\left( 
\begin{array}{cccccc}
k_{1,1} & k_{1,2} & k_{1,3} & k_{1,4} &  &  \\ 
k_{2,1} & k_{2,2} & k_{2,3} & k_{2,4} &  &  \\ 
k_{3,1} & k_{3,2} & k_{3,3} & k_{3,4} &  &  \\ 
k_{4,1} & k_{4,2} & k_{4,3} & k_{4,4} &  &  \\ 
&  &  &  & k_{5,5} & k_{5,6} \\ 
&  &  &  & k_{6,5} & k_{6,6}%
\end{array}%
\right) 
\]%
where we identify $7$ {\rm Type-I} solutions%
\[
{\Bbb K}_{[12]}^{(0)}=\left( 
\begin{array}{cccccc}
k_{1,1} & k_{1,2} &  &  &  &  \\ 
k_{2,1} & k_{2,2} &  &  &  &  \\ 
&  & k_{3,3} &  &  &  \\ 
&  &  & k_{4,4} &  &  \\ 
&  &  &  & k_{5,5} &  \\ 
&  &  &  &  & k_{6,6}%
\end{array}%
\right) ,\cdots ,{\Bbb K}_{[56]}^{(0)}=\left( 
\begin{array}{cccccc}
k_{1,1} &  &  &  &  &  \\ 
& k_{2,2} &  &  &  &  \\ 
&  & k_{3,3} &  &  &  \\ 
&  &  & k_{4,4} &  &  \\ 
&  &  &  & k_{5,5} & k_{5,6} \\ 
&  &  &  & k_{6,5} & k_{6,6}%
\end{array}%
\right) ,
\]%
In addition to ${\Bbb K}_{[14]}^{(0)}$ we have the {\rm Type-II} solution%
\[
{\Bbb K}_{[14]}^{(1)}=\left( 
\begin{array}{cccccc}
k_{1,1} &  &  & k_{1,4} &  &  \\ 
& k_{2,2} & k_{2,3} &  &  &  \\ 
& k_{3,2} & k_{3,3} &  &  &  \\ 
k_{4,1} &  &  & k_{4,4} &  &  \\ 
&  &  &  & k_{5,5} &  \\ 
&  &  &  &  & k_{6,6}%
\end{array}%
\right) 
\]%
with the constraint equation $k_{1,4}k_{4,1}=k_{2,3}k_{3,2}$, (the pairs of
entries of the same secondary diagonal) and the {\rm Type-III} solution%
\[
{\Bbb K}_{[14][56]}^{(1)\ (0)}=\left( 
\begin{array}{cccccc}
k_{1,1} &  &  & k_{1,4} &  &  \\ 
& k_{2,2} & k_{2,3} &  &  &  \\ 
& k_{3,2} & k_{3,3} &  &  &  \\ 
k_{4,1} &  &  & k_{4,4} &  &  \\ 
&  &  &  & k_{5,5} & k_{5,6} \\ 
&  &  &  & k_{6,5} & k_{6,6}%
\end{array}%
\right) 
\]%
with the constraint equation $k_{1,4}k_{4,1}=k_{2,3}k_{3,2}=k_{5,6}k_{6,5}$
since $(1+4)=(5+6)$ $\mod{6}$, $k_{1,4}\in K^{b}$ and $k_{5,6}\in K^{f}$.

Although we already know as counting the $K$-matrices for the $%
U_{q}[sl(m|n)^{(1)}]$ models we still have to identify among them which are
similar. Indeed we can see a ${\Bbb Z}_{N}$ similarity transformation which
maps their matrix elements positions:%
\begin{equation}
K_{a}=g_{a}K_{0}g_{N-a},\quad a=0,1,2,\cdots ,N-1  \label{sol.6}
\end{equation}%
where $g_{a}$ are the ${\Bbb Z}_{N}$ matrices%
\begin{equation}
\left( g_{a}\right) _{i,j}=\delta _{i,i+a}\qquad {\rm mod\ }N  \label{sol.7}
\end{equation}%
In order to do this we can choose $K_{0}$ as ${\Bbb K}_{[12]}^{(\alpha )}$
and the similarity transformations (\ref{sol.6}) give us the $K_{a}$
matrices whose matrix elements are in the same positions of the matrix
elements of the ${\Bbb K}_{[1j]}^{(\alpha )}$ and ${\Bbb K}_{[2m]}^{(\alpha
)}$ matrices. However, due to the fact that the relations (\ref{sol.2})
involve the ratio $c_{2}(x)/c_{1}(x)=x$, as well as the additional
constraints (\ref{rule2}), we could not find a similarity transformation
among these ${\Bbb K}^{^{\prime }}s$ matrices, even after a gauge
transformation. Even for the {\rm Type-I} solutions the similarity account
is not simple due to the presence of three types of scalar functions and the
constraint equations for the parameters $\beta _{i,j}$. Nevertheless, as we
have found a way to write all solutions, we can leave the similarity account
to the reader.

Having identified these possibilities we may proceed in order to find the $N$
diagonal elements $k_{i,i}(x)$ in terms of the non-diagonal elements $%
k_{i,j}(x)$ for each ${\Bbb K}_{i,j}^{(\alpha )}$ matrix. These procedure is
now standard \cite{Lima1}. For instance, if we are looking for ${\Bbb K}%
_{[14][56]}^{(1)\ (0)}$, the non-diagonal elements $k_{i,j}(x),$ ($i+j=5%
\mod{6}$ ) \ in terms of $k_{1,4}(x)\neq 0$ are given by%
\begin{eqnarray}
k_{2,3}(x) &=&\frac{\beta _{2,3}}{\beta _{1,4}}k_{1,4}(x),\quad k_{3,2}(x)=%
\frac{\beta _{3,2}}{\beta _{1,4}}k_{1,4}(x),\quad k_{4,1}(x)=\frac{\beta
_{4,1}}{\beta _{1,4}}k_{1,4}(x),  \nonumber \\
k_{5,6}(x) &=&\frac{\beta _{5,6}}{\beta _{1,4}}xk_{1,4}(x),\quad k_{6,5}(x)=%
\frac{\beta _{6,5}}{\beta _{1,4}}xk_{1,4}(x).  \label{sol.8}
\end{eqnarray}

Substituting (\ref{sol.8}) into the reflection equations we can now easily
find the $k_{i,i}(x)$ elements up to an arbitrary function, in this example
identified as $k_{1,4}(x)$. Moreover, their consistency relations will yield
us some constraints equations for the parameters $\beta _{i,j}$.

After we have found all diagonal elements in terms of $k_{i,j}(x)$, we can,
without loss of generality, choose the arbitrary functions as%
\begin{equation}
k_{i,j}(x)=\frac{1}{2}\beta _{i,j}(x^{2}-1),\qquad i<j.  \label{sol.9}
\end{equation}%
This choice allows us to work out the solutions in terms of the functions $%
f_{i,i}(x)$ and $h_{i,j}(x)$ defined by%
\begin{equation}
f_{ii}(x)=\beta _{i,i}(x-1)+1\qquad {\rm and}\qquad h_{ij}(x)=\frac{1}{2}%
\beta _{i,j}(x^{2}-1),  \label{sol.10}
\end{equation}%
for $i,j=1,2,\cdots ,N.$

Now, we will simply present the general solutions and write them explicitly
for the first values of $N$ in appendices.

\subsection{The quasi-diagonal K-matrices}

\ For{\rm \ Type-I} and {\rm Type-II} solutions we have the same general $K$%
-matrix form 
\begin{eqnarray}
{\Bbb K}_{[i,j]}^{[\alpha ]} &=&\sum_{k=0}^{\alpha }\left\{ f_{ii}(x){\rm E}%
_{i+ki+k}+h_{i+kj-k}(x){\rm E}_{i+kj-k}+h_{j-ki+k}(x){\rm E}_{j-ki+k}\right. 
\nonumber \\
&&\left. +x^{2}f_{ii}(x^{-1}){\rm E}_{j-kj-k}\right\} +{\cal Z}%
_{i}(x)\sum_{l=1}^{i-1}{\rm E}_{ll}+{\cal Y}_{i+1+\alpha
}^{(i)}(x)\sum_{l=i+1+\alpha }^{j-1-\alpha }{\rm E}_{ll}  \nonumber \\
&&+(1-\delta _{1,i})\ x^{2}{\cal Z}_{i}(x)\sum_{l=j+1}^{N}{\rm E}%
_{ll}+\delta _{1,i}\ {\cal X}_{j+1}(x)\sum_{l=j+1}^{N}{\rm E}_{ll}  \nonumber
\\
1 &\leq &i<j\leq m\quad {\rm or}\quad m+1\leq i<j\leq m+n  \label{ksol1}
\end{eqnarray}%
For the{\rm \ Type-III} solutions we have matrices with non-diagonal entries
into two secondary diagonals with different degree of freedom but related by 
$Z_{N}$ symmetry%
\begin{eqnarray}
{\Bbb K}_{[ij]\ [rs]}^{[\alpha ]\ [\beta ]} &=&\sum_{k=0}^{\alpha }\left\{
f_{ii}(x){\rm E}_{i+ki+k}+h_{i+kj-k}(x){\rm E}_{i+kj-k}+h_{j-ki+k}(x){\rm E}%
_{j-ki+k}\right.   \nonumber \\
&&\left. +x^{2}f_{ii}(x^{-1}){\rm E}_{j-kj-k}\right\} +{\cal Y}_{i+1+\alpha
}^{(i)}(x)\sum_{l=i+1+\alpha }^{j-1-\alpha }{\rm E}_{ll}  \nonumber \\
&&+\sum_{k=0}^{\beta }\left\{ x^{2}f_{ii}(x^{-1}){\rm E}%
_{r+kr+k}+xh_{r+ks-k}(x){\rm E}_{r+ks-k}+xh_{s-kr+k}(x){\rm E}%
_{r-ks+k}\right.   \nonumber \\
&&\left. +x^{2}f_{ii}(x){\rm E}_{s-ks-k}\right\} +{\cal X}_{r+1+\alpha
}(x)\sum_{l=r+1+\alpha }^{s-1-\alpha }{\rm E}_{ll}  \nonumber \\
1 &\leq &i<j\leq m\quad {\rm and}\quad m+1\leq r<s\leq m+n  \nonumber \\
i+j &=&r+s\quad \mod{N}  \label{ksol2}
\end{eqnarray}%
Note that for $\alpha ,\beta \neq 0$ we can use $\alpha =[\frac{j-i-1}{2}]$
and $\beta =[\frac{r-s-1}{2}]$. Moreover, we have defined more three types
of scalar functions 
\begin{eqnarray}
{\cal X}_{j+1}(x) &=&f_{11}(x^{-1})+\frac{1}{2}\left( \beta _{j+1,j+1}+\beta
_{1,1}-2\right) x^{-1}\left( x^{2}-1\right) ,  \nonumber \\
{\cal Y}_{l}^{(i)}(x) &=&f_{ii}(x)+\frac{1}{2}\left( \beta _{l,l}-\beta
_{i,i}\right) \left( x^{2}-1\right) ,  \nonumber \\
{\cal Z}_{i}(x) &=&f_{ii}(x^{-1})+\frac{1}{2}\left( \beta _{i,i}+\beta
_{1,1}\right) x^{-1}\left( x^{2}-1\right) .  \label{func}
\end{eqnarray}%
The number of free parameters is fixed by the constraint equations which
depend on the presence of these scalar functions: when ${\cal Y}_{l}^{(i)}(x)
$ is present in the $K$-matrix we have constraint equations of the type 
\begin{equation}
\beta _{i,j}\beta _{j,i}=\left( \beta _{l,l}+\beta _{i,i}-2\right) \left(
\beta _{l,l}-\beta _{i,i}\right) ,  \label{cstr1}
\end{equation}%
but, when ${\cal Z}_{i}(x)$ is present the corresponding constraints are of
the type%
\begin{equation}
\beta _{i,j}\beta _{j,i}=\left( \beta _{1,1}+\beta _{i,i}\right) \left(
\beta _{1,1}-\beta _{i,i}\right) .  \label{cstr2}
\end{equation}%
The presence of at least one ${\cal X}_{j+1}(x)$ yields a third type of
constraints, 
\begin{equation}
\beta _{i,j}\beta _{j,i}=\left( \beta _{j+1,j+1}+\beta _{1,1}-2\right)
\left( \beta _{j+1,j+1}-\beta _{1,1}-2\right) \text{.}  \label{cstr3}
\end{equation}%
Here we recall again that $i+j=r+s$ $\mod{N}$.

From (\ref{ksol1}) and (\ref{ksol2}) we can see that each solution we have
at most two scalar functions in addition to the $f_{ii}(x)$ and $\alpha
+\beta $ pairs of the $h(x)$ functions in addition to $h_{ij}(x)$ and $%
h_{rs}(x)$ functions. It means that our {\rm Type-I} matrices are $3$%
-parameter solutions. The {\rm Type-II} matrices have $3+\alpha $ free
parameters and the {\rm Type-III} matrices have $4+\alpha +\beta $ free
parameters.

\subsection{The diagonal K-matrices}

For diagonal solution we have $\beta _{i,j}=0$. It means that all scalar
functions $h_{i,j}(x)$ are equal to zero and we have to solve the constraint
equations (\ref{cstr1})-(\ref{cstr3}). Now, we can recall (\ref{ksol1}) and (%
\ref{ksol2}) \ and replace the scalar function ${\cal X}_{j+1}(x)$ by $%
x^{2}f_{11}(x^{-1})$ or by $x^{2}f_{11}(x)$ , the scalar function ${\cal Y}%
_{l}^{(i)}(x)$ by $f_{ii}(x)$ or by $x^{2}f_{ii}(x^{-1})$ and the scalar
function ${\cal Z}_{i}(x)$ by $f_{ii}(x^{-1})$ or by $f_{ii}(x)$ in order to
get the diagonal solutions. It follows due to the substitution of the
solutions of (\ref{cstr1})-(\ref{cstr3}) into (\ref{func}) 
\begin{eqnarray}
\lim_{\beta _{j,j}\rightarrow \pm \beta _{1,1}+2}{\cal X}_{j}(x)
&=&x^{2}f_{11}(x^{\mp 1})  \nonumber \\
\lim_{\beta _{l,l}\rightarrow \beta _{i,i}}{\cal Y}_{l}^{(i)}(x)
&=&f_{ii}(x)\qquad {\rm and}\qquad \lim_{\beta _{l,l}\rightarrow -\beta
_{i,i+2}}{\cal Y}_{l}^{(i)}(x)=x^{2}f_{ii}(x^{-1})  \nonumber \\
\lim_{\beta _{1,1}\rightarrow \pm \beta _{i,i}}{\cal Z}_{i}(x)
&=&f_{ii}(x^{\pm 1})  \label{lfunc}
\end{eqnarray}%
This reduction procedure gives us the diagonal solutions:

\begin{eqnarray}
{\Bbb D}_{[ij]} &=&{\sl Z}_{i}(x)\sum_{l=1}^{i-1}{\rm E}_{ll}+f_{ii}(x){\rm E%
}_{ii}+{\sl Y}_{i+1}^{(i)}(x)\sum_{l=i+1}^{j-1}{\rm E}%
_{ll}+x^{2}f_{ii}(x^{-1}){\rm E}_{jj}  \nonumber \\
&&+(1-\delta _{1,i})\ x^{2}{\sl Z}_{i}(x)\sum_{l=j+1}^{N}{\rm E}_{ll}+\delta
_{1,i}\ {\sl X}_{j+1}(x)\sum_{l=j+1}^{N}{\rm E}_{ll}  \nonumber \\
1 &\leq &i\,<j\leq m\quad {\rm or}\quad m+1\leq i\,<j\leq m+n  \label{kdiag1}
\end{eqnarray}%
and%
\begin{eqnarray}
{\Bbb D}_{[ij]\ [rs]} &=&f_{ii}(x){\rm E}_{ii}+x^{2}f_{ii}(x^{-1}){\rm E}%
_{jj}+{\sl Y}_{i+1}^{(i)}(x)\sum_{l=i+1}^{j-1}{\rm E}_{ll}  \nonumber \\
&&+x^{2}f_{ii}(x^{-1}){\rm E}_{rr}+x^{2}f_{ii}(x){\rm E}_{ss}+{\sl X}%
_{r+1}(x)\sum_{l=r+1}^{s-1}{\rm E}_{ll}  \nonumber \\
1 &\leq &i<j\leq m\quad {\rm and}\quad m+1\leq r<s\leq n+m  \label{kdiag2}
\end{eqnarray}%
where%
\begin{eqnarray}
{\sl X}_{j+1}(x) &=&\{x^{2}f_{11}(x^{-1}),x^{2}f_{11}(x)\},  \nonumber \\
{\sl Y}_{i+1}^{(i)}(x) &=&\left\{ x^{2}f_{ii}(x^{-1}),f_{ii}(x)\right\} , 
\nonumber \\
{\sl Z}_{i}(x) &=&\{f_{ii}(x^{-1}),f_{ii}(x)\}.  \label{dfunc}
\end{eqnarray}

From these results we can see that the $U_{q}[sl(m|n)^{(1)}]$ model have
many diagonal solutions. In particular, the substitution ${\sl Z}%
_{i}(x)=f_{ii}(x)$, ${\sl Y}_{i+1}^{(i)}(x)=f_{ii}(x)$ and ${\sl X}%
_{j+1}=x^{2}f_{11}(x^{-1})$ into (\ref{kdiag1}) yields the diagonal
solutions already derived in \cite{Yue1} and used in the study of the nested
Bethe ansatz for Perk-Scultz model with open boundary condition \cite{Yue2}.
Moreover, \ these diagonal solutions have been used recently in \cite{BR}
for the study of the nested Bethe ansatz for 'all' open chain with diagonal
boundary conditions.

\section{Conclusion}

After a systematic study of the functional equations we find that there are
three types of solutions for $U_{q}[sl(m|n)^{(1)}]$ model. We call of {\rm %
Type-I} the $K$-matrices with three free parameters and $n+m+2$ non-zero
matrix elements. These solutions were denoted by ${\Bbb K}_{[ij]}^{(0)}$ to
emphasize the non-zero element out of the diagonal and its symmetric, which
results in $n(n-1)/2$ and $m(m-1)/2$ reflection $K$-matrices.

The {\rm Type-II} and {\rm Type-III} solutions are more interesting because
their have many free parameters. We also have used a reduction procedure to
obtain the diagonal solutions. However, we could not derive a similar
procedure in order to obtain the {\rm Type-I} solutions from the {\rm Type-II%
} solutions or the {\rm Type-II} solutions from \ the{\rm Type-III}
solutions. Thus, we believe that they are independent.

The corresponding $K^{+}(x)$ are obtained from the isomorphism (\ref{re.8}).
Out of this classification we have the trivial solution $\left(
K^{-}=1,K^{+}=M\right) $ for these models. 

Before the end of our discussion on the $U_{q}[sl(m|n)^{(1)}]$ reflection
matrices, we will make (by a referee suggestion), the comparision with the $%
sl(m+n)$ reflection matrices \cite{Lima1}. The diagonal solutions and the 
{\rm Type-I} solutions are the same for the both models. The {\rm Type-III}
solutions of the $U_{q}[sl(m|n)^{(1)}]$ model are identified with the {\rm %
Type-II} of the $sl(m+n)$ model. However, the {\rm Type-II} solutions of the 
$U_{q}[sl(m|n)^{(1)}]$ model are different because in the graded case, the $%
Z_{N}$ symmetry (\ref{rule3}) is lost when the labels have the same degree
of freedom (\ref{rule2}).

{\bf Acknowledgment:} This work was supported in part by Funda\c{c}\~{a}o de
Amparo \`{a} Pesquisa do Estado de S\~{a}o Paulo--FAPESP--Brasil and by
Conselho Nacional de Desenvol\-{}vimento--CNPq--Brasil.

\appendix

\section{Some examples}

In this appendix some K-matrices are written explicitly only for the cases
with $m\geq n$. \ The cases $m\,<n$ are easily deduced from the $%
U_{q}[sl(m|n)^{(1)}]$ solutions with $m>n$ using (\ref{sol.6}).

\bigskip For the $U_{q}[sl(1|1)^{(1)}]$ model there is only one diagonal $K$%
-matrix \ 
\begin{equation}
{\Bbb D}_{[12]}=\left( 
\begin{array}{cc}
f_{11}(x) & 0 \\ 
0 & x^{2}f_{11}(x^{-1})%
\end{array}%
\right)  \label{a.1}
\end{equation}

It follows from (\ref{ksol1}) that we have only one {\rm Type-I} solution $%
{\Bbb K}_{[12]}^{(0)}$ for $U_{q}[sl(2|1)^{(1)}]$ model : 
\begin{eqnarray}
{\Bbb K}_{[12]}^{(0)} &=&f_{11}(x){\rm E}_{11}+h_{12}(x){\rm E}%
_{12}+h_{21}(x){\rm E}_{21}+x^{2}f_{11}(x^{-1}){\rm E}_{22}+{\cal X}_{3}(x)%
{\rm E}_{33}  \nonumber \\
&=&\left( 
\begin{array}{ccc}
f_{11}(x) & h_{12}(x) & 0 \\ 
h_{21}(x) & x^{2}f_{11}(x^{-1}) & 0 \\ 
0 & 0 & {\cal X}_{3}(x)%
\end{array}%
\right) ,  \label{a.2}
\end{eqnarray}%
where four parameters $\beta _{11},\beta _{12},\beta _{21}$ and $\beta _{33}$
satisfied the constraint equation%
\begin{equation}
\beta _{12}\beta _{21}=\left( \beta _{33}-\beta _{11}-2\right) \left( \beta
_{33}+\beta _{11}-2\right) .  \label{a.3}
\end{equation}%
Two diagonal solutions are derived from (\ref{a.2}) due to the constraint
equation (\ref{a.3})%
\begin{eqnarray}
\lim_{\beta _{33}\rightarrow \mp \beta _{11}+2}{\cal X}_{3}(x)
&=&x^{2}f_{11}(x^{\mp 1})  \nonumber \\
&\Rightarrow &{\Bbb D}_{[12]a}={\rm diag}%
(f_{11}(x),x^{2}f_{11}(x^{-1}),x^{2}f_{11}(x^{-1}))  \nonumber \\
&\Rightarrow &{\Bbb D}_{[12]b}={\rm diag}%
(f_{11}(x),x^{2}f_{11}(x^{-1}),x^{2}f_{11}(x))
\end{eqnarray}

The solutions ${\Bbb D}_{[12]a}$ is the diagonal solution derived by the
first time in \cite{Yue1}.

For $U_{q}[sl(3|1)^{(1)}]$ \ model we have three {\rm Type-I} matrices:%
\begin{eqnarray}
{\Bbb K}_{[12]}^{(0)} &=&\left( 
\begin{array}{cccc}
f_{11}(x) & h_{12}(x) & 0 & 0 \\ 
h_{21}(x) & x^{2}f_{11}(x^{-1}) & 0 & 0 \\ 
0 & 0 & {\cal X}_{3}(x) & 0 \\ 
0 & 0 & 0 & {\cal X}_{3}(x)%
\end{array}%
\right)  \nonumber \\
\beta _{12}\beta _{21} &=&\left( \beta _{33}+\beta _{11}-2\right) \left(
\beta _{33}-\beta _{11}-2\right) ,
\end{eqnarray}%
with two diagonals%
\begin{eqnarray}
{\Bbb D}_{[12]a} &=&{\rm diag}%
(f(x),x^{2}f(x^{-1}),x^{2}f(x^{-1}),x^{2}f(x^{-1})),  \nonumber \\
{\Bbb D}_{[12]b} &=&{\rm diag}(f(x),x^{2}f(x^{-1}),x^{2}f(x),x^{2}f(x)),
\end{eqnarray}%
where $f(x)=\beta (x-1)+1$ and $\beta $ a free parameter,%
\begin{eqnarray}
{\Bbb K}_{13}^{(0)} &=&\left( 
\begin{array}{cccc}
f_{11}(x) & 0 & h_{13}(x) & 0 \\ 
0 & {\cal Y}_{2}^{(1)}(x) & 0 & 0 \\ 
h_{31}(x) & 0 & x^{2}f_{11}(x^{-1}) & 0 \\ 
0 & 0 & 0 & {\cal X}_{4}(x)%
\end{array}%
\right) \\
\beta _{13}\beta _{31} &=&\left( \beta _{44}+\beta _{11}-2\right) \left(
\beta _{44}-\beta _{11}-2\right) =\left( \beta _{22}+\beta _{11}-2\right)
\left( \beta _{22}-\beta _{11}\right) ,
\end{eqnarray}%
with four diagonals%
\begin{eqnarray}
{\Bbb D}_{[13]a} &=&{\Bbb D}_{[12]a},  \nonumber \\
{\Bbb D}_{[13]b} &=&{\rm diag}(f(x),f(x),x^{2}f(x^{-1}),x^{2}f(x^{-1})), 
\nonumber \\
{\Bbb D}_{[13]c} &=&{\rm diag}(f(x),x^{2}f(x^{-1}),x^{2}f(x^{-1}),x^{2}f(x)),
\nonumber \\
{\Bbb D}_{[13]d} &=&{\rm diag}(f(x),f(x),x^{2}f(x^{-1}),x^{2}f(x)).
\end{eqnarray}%
and%
\begin{eqnarray}
{\Bbb K}_{[23]}^{(0)} &=&\left( 
\begin{array}{cccc}
{\cal Z}_{2}(x) & 0 & 0 & 0 \\ 
0 & f_{22}(x) & h_{23}(x) & 0 \\ 
0 & h_{32}(x) & x^{2}f_{22}(x^{-1}) & 0 \\ 
0 & 0 & 0 & x^{2}{\cal Z}_{2}(x)%
\end{array}%
\right)  \nonumber \\
\beta _{23}\beta _{32} &=&\left( \beta _{11}+\beta _{22}\right) \left( \beta
_{11}-\beta _{22}\right) ,
\end{eqnarray}%
with two diagonals%
\begin{eqnarray}
{\Bbb D}_{[23]a} &=&{\Bbb D}_{[13]d}, \\
{\Bbb D}_{[23]b} &=&{\rm diag}(f(x^{-1}),f(x),x^{2}f(x^{-1}),x^{2}f(x^{-1}))
\nonumber
\end{eqnarray}

For $U_{q}[sl(2|2)^{(1)}]$ model we have the same {\rm Type-I} \ ${\Bbb K}%
_{[34]}^{(0)}$\ , ${\Bbb K}_{[12]}^{(0)}$ matrices written above and one 
{\rm Type-III} matrix with four free parameters $\beta _{1,2}$, $\beta
_{2,1} $, $\beta _{3,4}$ and $\beta _{1,1}$:%
\begin{eqnarray}
{\Bbb K}_{[12][34]}^{(0)\ (0)} &=&\left( 
\begin{array}{cccc}
f_{11}(x) & h_{12}(x) & 0 & 0 \\ 
h_{21}(x) & x^{2}f_{11}(x^{-1}) & 0 & 0 \\ 
0 & 0 & x^{2}f_{11}(x^{-1}) & xh_{34}(x) \\ 
0 & 0 & xh_{43}(x) & x^{2}f_{11}(x)%
\end{array}%
\right)  \nonumber \\
\beta _{12}\beta _{21} &=&\beta _{34}\beta _{43}.
\end{eqnarray}%
with one diagonal equal to ${\Bbb D}_{[13]c}.$ The diagonals with only
entries of the types $f(x)$ and $x^{2}f(x^{-1})$ are the solutions obtained
in \cite{Yue1}.

\end{document}